\documentclass{article}

\PassOptionsToPackage{numbers, compress}{natbib}
    \usepackage[final]{neurips_2022}

\usepackage[english]{babel}
\usepackage{tablefootnote}
\usepackage[caption = false]{subfig}
\usepackage{graphicx}
\usepackage{float}
\usepackage[utf8]{inputenc} 
\usepackage[T1]{fontenc}    
\usepackage{hyperref}       
\usepackage{url}            
\usepackage{booktabs}       
\usepackage{amsfonts}       
\usepackage{amsmath}
\usepackage{nicefrac}       
\usepackage{microtype}      
\usepackage{xcolor}         

\title{From Competition to Collaboration: Making Toy Datasets on Kaggle Clinically Useful for Chest X-Ray Diagnosis Using Federated Learning}

\author{%
  Pranav Kulkarni$^{1}$ \quad Adway Kanhere$^{1,2}$ \quad Paul H. Yi$^{1}$ \quad Vishwa S. Parekh$^{1}$\\
  $^{1}$University of Maryland Medical Intelligent Imaging (UM2ii) Center \\
  Department of Diagnostic Radiology and Nuclear Medicine \\
  University of Maryland School of Medicine \\
  Baltimore, MD 21201 \\
  \texttt{\{pkulkarni, pyi, vparekh\}@som.umaryland.edu} \\
  $^{2}$Department of Biomedical Engineering \\
  Johns Hopkins University \\
  Baltimore, MD 21218 \\
  \texttt{akanher1@jhu.edu}
}

\begin{document}

\maketitle

\begin{abstract}
    Chest X-ray (CXR) datasets hosted on Kaggle, though useful from a data science competition standpoint, have limited utility in clinical use because of their narrow focus on diagnosing one specific disease. In real-world clinical use, multiple diseases need to be considered since they can co-exist in the same patient. In this work, we demonstrate how federated learning (FL) can be used to make these toy CXR datasets from Kaggle clinically useful. Specifically, we train a single FL classification model (`global`) using two separate CXR datasets -- one annotated for presence of pneumonia and the other for presence of pneumothorax (two common and life-threatening conditions) -- capable of diagnosing both. We compare the performance of the global FL model with models trained separately on both datasets (`baseline`) for two different model architectures. On a standard, naive 3-layer CNN architecture, the global FL model achieved AUROC of 0.84 and 0.81 for pneumonia and pneumothorax, respectively, compared to 0.85 and 0.82, respectively, for both baseline models (p>0.05). Similarly, on a pretrained DenseNet121 architecture, the global FL model achieved AUROC of 0.88 and 0.91 for pneumonia and pneumothorax, respectively, compared to 0.89 and 0.91, respectively, for both baseline models (p>0.05). Our results suggest that FL can be used to create global `meta` models to make toy datasets from Kaggle clinically useful, a step forward towards bridging the gap from bench to bedside.
\end{abstract}

\section{Introduction}

Chest X-Ray (CXR) is the most commonly ordered medical imaging study globally and is critical for screening many life threatening conditions (e.g., pneumonia). Accordingly, many large-scale public CXR datasets have been released through curation of  \cite{wang2017chestx,irvin2019chexpert,johnson2019mimic,nguyen2022vindr}. These, in turn, have resulted in numerous data science competitions hosted on platforms like Kaggle (e.g., RSNA pneumonia detection challenge \cite{shih2019augmenting}), resulting in expert-level performance for disease diagnoses.

Although useful from a data science competition standpoint, these Kaggle-hosted CXR datasets have limited clinical utility because of their narrow focus on one single  diagnostic task. For example, the two Kaggle CXR competitions hosted by RSNA and SIIM have focused on diagnosis of a single disease, like pneumonia and pneumothorax \cite{yi2021demographic}. Although impressive results have resulted from these competitions, their utility is limited given the dozens of diagnoses that could present in real-world clinical practice. Therefore, a method to harmonize these toy datasets to train a clinically-useful model could revolutionize how small, narrowly-focused datasets can be leveraged in aggregate for development of clinically-relevant deep learning models.

We propose CheXViz, a federated learning (FL) framework for training a single model on spatially distributed datasets with different disease annotations into a `global` meta-deep learning model. Briefly, FL is a machine learning technique that approaches the problem from a multi-domain and multi-task perspective. By using a decentralized and distributed approach, consisting of a central server and nodes, a global `meta` model can be trained to generalize distributed tasks with non-iid labels. During each training step (`FL round`), every node trains a local model. Then, the weights across all nodes are aggregated by the central server and redistributed back to the nodes. In medical imaging, FL has enabled training of large-scale `global` deep learning models using datasets spread across multiple institutions without sharing sensitive patient data. In this preliminary work, we demonstrate the utility of CheXViz for training a single model to diagnose pneumonia and pneumothorax using two toy datasets from Kaggle for these two respective diseases. Put another way, we demonstrate how toy datasets from Kaggle can be made clinically useful using FL.
\begin{figure}[H]
    \centering
    \includegraphics[width=1\textwidth]{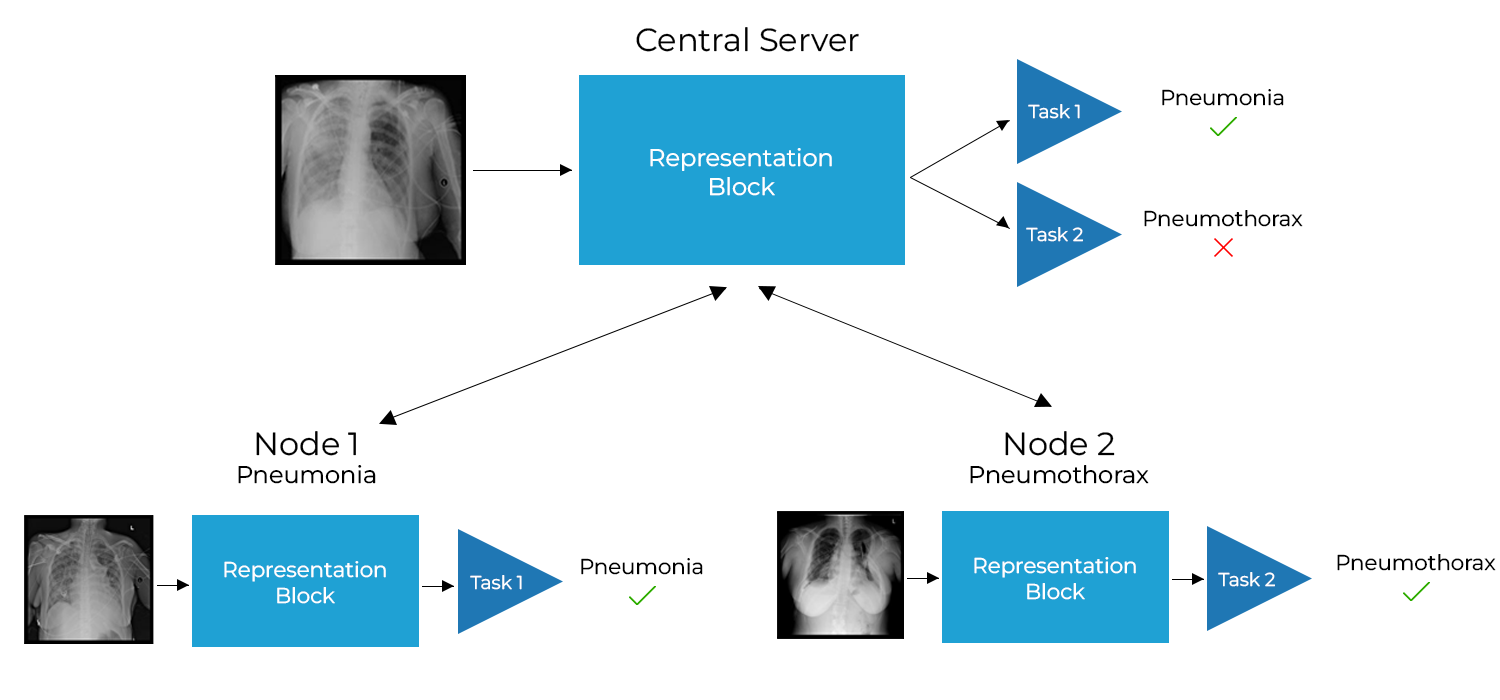}
    \caption{Illustration of the CheXViz framework}
    \label{fig:fl}
\end{figure}

\section{Methods}
\subsection{CheXViz}
We developed CheXViz using a multi-task FL setup. The CheXViz model is initialized as a deep neural network consisting of two distinct blocks - a representation block and a task block. The CheXViz model is distributed across all the participating nodes to train their tasks. During training, only the weights corresponding to the representation block are aggregated and redistributed by the central server back to the nodes, thereby preserving task-related information for each node in their task block. Figure \ref{fig:fl} illustrates the CheXViz framework. 

\subsection{Experimental Setup}
We evaluated the CheXViz framework for the task of training a generalized global model that can classify cases of pneumonia and pneumothorax using distributed and non-iid CXR datasets from the RSNA Pneumonia Detection and the SIIM-ACR Pneumothorax Segmentation competitions on Kaggle. Our experiments involved two different deep network architectures: A naive convolutional neural network (CNN) consisting of 3 convolutional layers (`standard` model) and a pre-trained DenseNet121 architecture using transfer learning (TL) \cite{densenet}. We implemented Federated Averaging (FedAvg) for model weight aggregation as described in \cite{fedavg}. 

The standard baseline models for pneumonia and pneumothorax classification were trained for 300 epochs using a learning rate scheduler with an initial learning rate of 1e-3. During FL, we trained the standard global model for 300 FL rounds where for each round, models were locally trained for 1 epoch before aggregation. The DenseNet121 model was initialized with pre-trained ImageNet weights (`base model`) and a new classification head was trained for 30 epochs with a learning rate of 1e-3 while the rest of the base model weights were frozen and then fine tuned with a slower learning rate of 1e-5 to prevent overfitting. For the baseline models, the models were fine tuned for 150 epochs. For FL, we utilized FedAvg during the fine tuning step of TL. The global model was fine tuned for 150 FL rounds with 1 epoch before aggregation with a learning rate of 1e-5. All the models were trained and evaluated on a Google Cloud VM with four NVIDIA T4 GPUs.

We computed sensitivity, specificity, area under precision-recall (AUPR) curve, and the area under receiver operating characteristic (AUROC) curve for each model. The AUROC values between the FL global model and the task's baseline model were compared using bootstrapping and a paired t-test. Statistical significance was defined as $p < 0.05$.

\section{Results}

\begin{figure}[!htb]
    \centering
    \subfloat{\includegraphics[width = 2in]{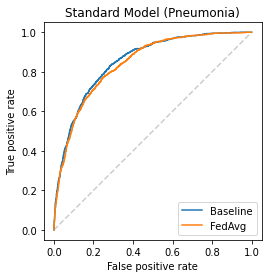}} 
    \subfloat{\includegraphics[width = 2in]{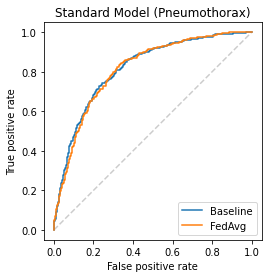}}\\
    \subfloat{\includegraphics[width = 2in]{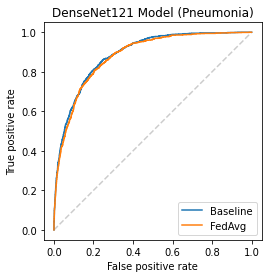}}
    \subfloat{\includegraphics[width = 2in]{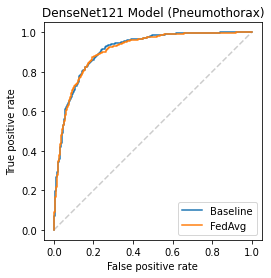}} 
    \caption{ROC curves obtained from  baseline and CheXViz models evaluated across both the datasets.}
    \label{fig:roc}
\end{figure}

The CheXViz framework trained `meta` models demonstrated excellent performance compared to the baseline models for the diagnostic classification of pneumonia and pneumothorax abnormalities, as shown in Figure \ref{fig:roc}. For pneumonia classification with the standard model architecture, the baseline and FL models achieved a validation AUROC of 0.85 and 0.84 respectively. We observed a similar trend with the DenseNet121 architecture, where the baseline and FL models achieved a validation AUROC of 0.89 and 0.88 respectively. For pneumothorax classification with the standard model architecture, the baseline and FL models achieved a validation AUROC of 0.82 and 0.81 respectively. Again, a similar result was achieved with the DenseNet121 architecture, where the baseline and FL models both achieved a validation AUROC of 0.91. The results are detailed in Appendix Table \ref{tab:metrics}. 

We further visualized the Grad-CAM outputs for evaluating the explainability and generalizability of the models \cite{gradcam}. Our preliminary analysis suggests that the heatmaps from CheXViz models demonstrate higher and focused activations within the lungs, compared to the baseline models as shown in Appendix Figure \ref{fig:gradcam}. For future work, we intend to outline a methodology to quantify the generalizability of models for CXR classification using Grad-CAM heatmaps.

\section{Discussion}

Although Kaggle CXR datasets and data science competitions have made an indelible impact on data science and AI for healthcare, they are still a far cry from being clinically useful datasets. This is understandable, given the challenges in curating expert-level annotations for diseases, and ostensibly why these Kaggle-hosted competitions have focused largely on single diseases\cite{shih2019augmenting,yi2021demographic}. Nevertheless, there is a gap between these toy datasets and clinical utility. Put another way, it is unclear how to use these datasets to train clinically-useful models capable of detecting multiple diseases. Our findings demonstrate that our FL framework (CheXViz) can be used to create global `meta` models to make toy datasets from Kaggle clinically useful, a large step forward towards bridging the gap from bench to bedside. Although preliminary in nature and focusing on only two datasets, our framework and results are extensible to any number of datasets and disease labels, as well as tasks beyond classification (e.g., segmentation and object detection). It is our hope that our work can be a first step towards moving Kaggle CXR datasets from competition to collaboration and transform these toy datasets into clinically useful models.

\bibliography{main}

\section*{Societal Impact}
Our work has a potential positive societal impact by taking an important step towards translational research. Our work can be a first step towards moving Kaggle CXR datasets from competition to collaboration and transform these toy datasets into clinically useful models. However, the use of FL can potentially lead to privacy and security risks, making the system vulnerable to client and server attacks, thereby leading to potentially negative societal impacts. We are actively working to address these vulnerabilities in FL systems.

\appendix

\section{Appendix}

\begin{table}[!htb]
  \centering
  \caption{Model Metrics}
  \begin{tabular}{llcccccc}
    \toprule
    \textbf{Task} & \textbf{Model} & \textbf{Loss} & \textbf{Sensitivity} & \textbf{Specificity} & \textbf{AUPR} & \textbf{AUROC} & \textbf{p-value} \\
    \midrule
    Pneumonia & Standard & 0.38 & 82.57 & 71.97 & 0.63 & 0.85 & - \\
    & Standard w/ FL & 0.39 & 78.01 & 74.41 & 0.61 & 0.84 & 0.10 \\
    \cmidrule(r){2-8}
    & DenseNet121 & 0.34 & 84.90 & 76.76 & 0.71 & 0.89 & - \\
    & DenseNet121 w/ FL & 0.35 & 80.08 & 79.79 & 0.70 & 0.88 & 0.19 \\
    \midrule
    Pneumothorax & Standard & 0.41 & 74.13 & 75.89 & 0.54 & 0.82 & - \\
    & Standard w/ FL & 0.42 & 80.22 & 69.87 & 0.52 & 0.81 & 0.71 \\
    \cmidrule(r){2-8}
    & DenseNet121 & 0.31 & 91.30 & 76.31 & 0.73 & 0.91 & - \\
    & DenseNet121 w/ FL & 0.31 & 84.57 & 83.10 & 0.73 & 0.91 & 0.76 \\
    \bottomrule
  \end{tabular}
  \label{tab:metrics}
\end{table}

\begin{figure}[!htb]
    \centering
    \subfloat[Pneumonia]{\includegraphics[width = 2.7in]{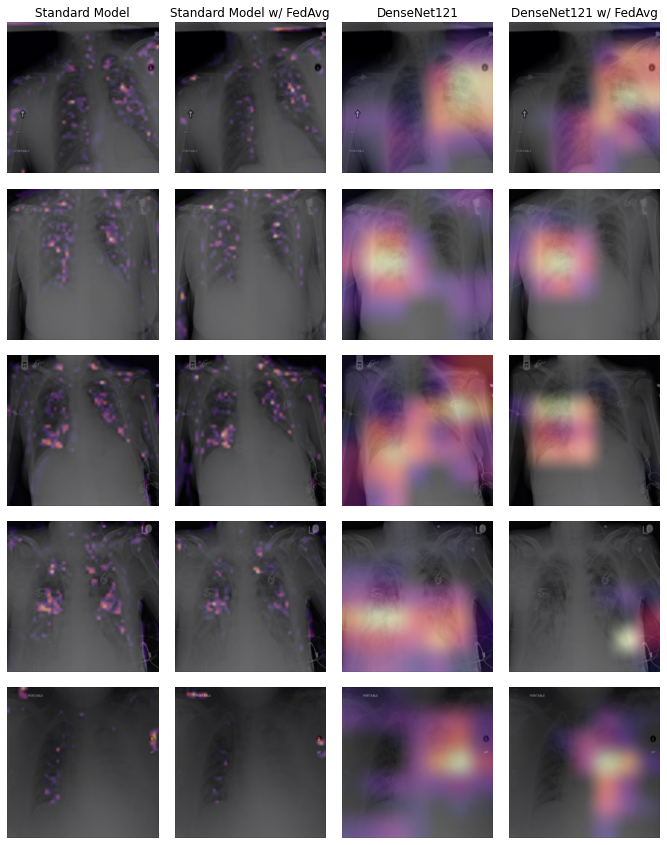}}
    \hspace{0.25em}
    \subfloat[Pneumothorax]{\includegraphics[width = 2.7in]{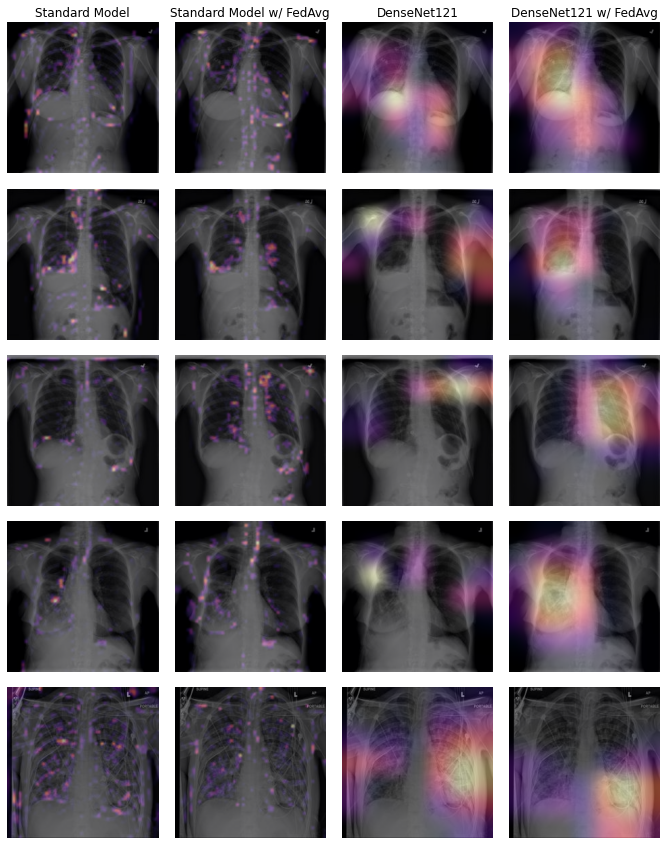}}\\
    \caption{Grad-CAM Visualization of True Positives}
    \label{fig:gradcam}
\end{figure}

\end{document}